\documentclass[a4paper,12pt]{article}
\usepackage{helvet}
\usepackage{amsthm}
\newtheorem{theorem}{Theorem}
\title{Sat Has No Wizards}
\author{Silvano Di Zenzo \\ Department of Computer Science, University of Rome}
\date{}
\begin{document}
\maketitle

\begin{abstract}
 An (encoded) decision problem over $\Sigma$ is a pair $(E, F)$ where $E$=words that encode instances of the problem, $F$=words to be accepted. We use $strings$ in a technical sense, borrowed from Computability. With any NP problem $(E, F)$ we associate a set of strings $|Log_E(F)|$ called the reduced logogram of $F$ relative to $E$, which conveys structural information on $E$, $F$, and how $F$ is embedded in $E$. We define notions of internal independence of decision problems in terms of $|Log_E(F)|$. The kernel $Ker(P)$ of a program $P$ that solves $(E, F)$ is the set of those strings in $|Log_E(F)|$ that are actually used by $P$ in making decisions. There are strict relations between $Ker(P)$ and the complexity of $P$.
	
We develop an application to $SAT$ that relies upon a property of strong internal independence of $SAT$. We show that $SAT$ cannot have in its reduced logogram certain strings that, when present, serve as collective certificates. As a consequence, all the programs that solve $SAT$ have the same kernel $Ker(P)=|Log_{CNF} (SAT)|$.
\end{abstract}

\section{Introduction}
We develop an application to $SAT$, to be positioned in current stream of interest in the structure of Boolean satisfiability \cite{kolaitis}. We use $strings$ in a technical sense, borrowed from Computability \cite{odifreddi}.

An $encoded$  $decision$ $problem$ over $\Sigma$ is a pair $(E, F)$ where $E$=words that encode instances of the problem, $F$=words to be accepted. On input $x$, a decision program $P$ for $(E, F)$ either $accepts$ $x$ (if $x$ is in $F$) or $rejects$ $x$ (if $x$ is in $E-F$) or else $discards$ $x$ (for $x$ outside $E$).

Our fundamental construct is a set of strings $Log_E (F)$ called $logogram$ of $F$ relative to $E$ that conveys structural information on $E$, $F$, and how $F$ is embedded in $E$. We mostly use the reduced version $|Log_E (F)|$, consisting of those strings in $Log_E (F)$ that do not include other strings in $Log_E (F)$. The $kernel$ $Ker(P)$ of a program $P$ that solves $(E, F)$ is the set of those strings in $|Log_E (F)|$ that are actually used by $P$ in making decisions. There are strict relationships between the composition in terms of strings of the kernel $Ker(P)$ of a program solving $(E, F)$ and the complexity of $P$.

Our application to $SAT$ uses a property of internal independence of a decision problem that we call ``strong internal independence.'' Think of a computation in which the result of any computation step does not change the results that are possible for subsequent steps. Internal independence is defined in terms of a relation of $entanglement$ $\sqsupseteq^E$ between sets of strings relative to reference set $E$.

Our main results are the following. We show that $(CNF, SAT)$ exhibits the strong internal independence property: Intuitively, no ``entanglement at distance" between strings in $|Log_{CNF} (SAT)|$ is possible. Besides, we show that problem $(CNF, SAT)$ cannot have, in its reduced logogram, certain collective certificates that we call $wizards$. As a consequence, the decision programs $P$ that solve $(CNF, SAT)$ all have the same kernel $Ker(P)=|Log_{CNF} (SAT)|$.

\section{Certificates of Membership as Strings}
We first recall notions regarding the certificates of membership in NP theory. As next step, we illustrate possible use of strings to represent certificates. We conclude the section reviewing basic algebraic properties of strings.

Let $G\subseteq \Sigma^*\times\Sigma^*$ so that $G$ is a relation on words over $\Sigma$. Let $Dom(G)$ and $Cod(G)$ be first and second projection of $G$. A relation $G$ which is both polynomial-time decidable and polynomially balanced is an NP relation. $L$ is in NP if and only if there exists an NP relation $G$ such that $L=Dom(G)$. We interchange problems with languages: $(E, F) \in NP$ and $F \in NP$ amount to the same.

Let $(E, F)$ be an NP problem. Then there exists a sequence $y_1$, $y_2$,.. of words (over some appropriate alphabet) called $solutions$ or else $certificates$ $of$ $membership$ for problem $(E, F)$. For any problem instance $x\in E$ we have that $x$ can possibly be satisfied by some of the $y_i s$. We also have ``unsatisfiable'' instances. What ``satisfaction'' means operationally is proper of problem $(E, F)$.

Cardinality function $\alpha (n)$ of an NP problem: We may arrange notations so that all solutions that can possibly satisfy an $x$ of size $n$ are between $y_1$ and $y_{\alpha (n)}$.

Associated with solutions $y_1$, $y_2$,.. there is a decomposition of target set $F$ into subsets $F_i$ called $regions$, where $F_i$ is the set of those $x's$ that are satisfied by $y_i$. Regions satisfy the obvious relation $F=\cup_i F_i$.

\subsection{Generalized Certificates}
In this paper we replace certificates with $generalized$ $certificates$. These are represented by $strings$, defined to be functions $N\rightarrow\Sigma$ with finite domain ($N$=positive integers). In loose words, a string $g$ being included (or subsumed) in a word $x$ is that which remains by canceling zero or more letters in $x$, while leaving blanks in places of letters. Note that words are certain special strings, thus the solutions $y_1$, $y_2$,.. continue to be certificates. This generalization allows us to introduce certain more general certificates that we call $wizards$.

We assume that satisfiability, being a property exhibited by certain words, is accompanied by characteristic $signs$, that we think as distinctive marks, or signatures, being somehow inscribed within the word $x$ under study. A detailed discussion would yield strings as proper formalization of such notions as ``mark'' or ``signature.'' Thus, we assume that signs are strings interspersed in $x$. Since strings represent words in shorthand, we call their set a $logogram$.

\subsection{Strings}
We define $\Sigma_\infty$ to be the set of all strings over $\Sigma$. Look at $g\in \Sigma_\infty$ as a prescription that a word $x$ over $\Sigma$ may or may not satisfy. If $Dom(g)$ is an initial segment of $N$ then $g$ is an ordinary word: Thus, words are certain special strings. The length (or size) $|g|$ is the greatest number in $Dom(g)$.

$\Sigma_\infty$ is partially ordered. Given $f, g \in \Sigma_\infty$, $g$ is an $extension$ of $f$, written $f \le g$, as soon as $Dom(f ) \subseteq Dom(g)$ and $g$ takes same values as $f$ in $Dom(f )$. If $f \le g$ and $g \le f$ then $f = g$. If $f \le g$ but not $g \le f$, write $f < g$ and say $g$ is a proper extension of $f$, or else $f$ is a proper $restriction$ of $g$. The empty partial function $N\rightarrow \Sigma$ , noted $\perp$, is the $void$ string, and  $Dom(\perp)=\emptyset$. Any $f$ in $\Sigma_\infty$ is an extension of $\perp$, thus $(\Sigma_\infty, \le)$ has a least element $\perp$.

Two strings $f$ and $g$ are $compatible$ as soon as $f(x)=g(x)$ for any $x$ in $Dom(f)\cap Dom(g)$. If $f, g$ are disjoint, which is to say $Dom(f)\cap Dom(g)=\emptyset$, then $f$ and $g$ are certainly compatible. The $meet$ $f \wedge g$ of any pair $f, g$ is the restriction of $f$ (or $g$) to that portion of the intersection $Dom(f)\cap Dom(g)$ where $f$ and $g$ agree. The $join$ of two compatible strings $f, g$, noted $f +g$, is the least string which is an extension of both $f$ and $g$. Thus $f,  g \le f+g$ and $Dom(f +g)=Dom(f ) \cup Dom(g)$. Equipped with meet and join, $\Sigma_\infty$ is an upward directed complete meet-semilattice \cite{gierz}.

\section{Entanglement among Strings}
The cylinders defined below are as in Computability (the formalism is slightly different). The logogram is a newcomer in Computer Science. Entanglement is a key concept to deal with internal structure of computational problems.

\subsection{Cylinders} Given $H \subseteq \Sigma_\infty$ we define
\begin{equation}\label{mauri1}
Exp(H)=\{x\in \Sigma^* : \exists a \in H \hspace{0.5em} (x \ge a) \}
\end{equation}
Thus, $Exp(H)$=set of all words that include strings from $H$. Call $Exp(H)$ $absolute$ $expansion$, equivalently, $absolute$ $cylinder$ associated with $H$. Note that $Exp(H)$ is the union of the elementary cylinders $Exp(g)$ for $g \in H$.

Given any recursive set of words $E$, we write $\Sigma_\infty (E)$ for the set of all strings that happen to be included in words of $E$, thus
\begin{equation}\label{mauri2}
\Sigma_\infty (E)=\{g\in\Sigma_\infty : Exp(g)\cap E \not= \emptyset \}
\end{equation}
$\Sigma_\infty (E)$ is the set of those strings $g$ in $\Sigma_\infty$  whose associated cylinder $Exp(g)$ intersects $E$. We think of $E$ as the set of words over $\Sigma$ that encode instances of some fixed reference computational problem $\Pi$. (Whenever we talk of a reference set $E$ there is implicit reference to some fixed abstract decision problem $\Pi$ as well as to a program $P$ solving $\Pi$.) For $H\subseteq \Sigma_\infty (E)$ we write
\begin{equation}\label{mauri3}
Exp_E (H)=E^H=\{x\in E : \exists a \in H \hspace{0.5em} (x \ge a) \}=E\cap Exp(H)
\end{equation}

Thus, $E^H$ is the set of those words in $E$ that contain strings from $H$. $E^H$ is the $expansion$ of $H$ relative to base $E$. We actually regard $E^H$ as a relativized cylinder, equivalently, as being a cylinder relative to a reference set $E$.
Note that for $E=\Sigma^*$ we regain the absolute expansion of set $H$.

Given $H \subseteq \Sigma_\infty (E)$, correspondence $Exp_E : H \rightarrow E^H$ exhibits properties:

\begin{equation}\label{mauri4}
E^H \cup E^K = E^{H \cup K}, \hspace{1em} E^H \cap E^K = E^{H + K}
\end{equation}

Thus, unions and intersections of sets that are cylinders relative to reference set $E$ are cylinders in $E$. Also note that, for any $H, K \in \Sigma_\infty (E)$,

\begin{equation}\label{mauri5}
H \subseteq K  \Rightarrow  E^H \subseteq E^K
\end{equation}
\begin{equation}\label{mauri6}
Exp_E (Exp_E(H))=Exp_E (E^H)=E^H
\end{equation}

\subsection{Logograms}
In this section we introduce the $logogram$ of a set of words $F$ relative to a reference set $E$. Given $F \subseteq E$, we define
\begin{equation}\label{mauri7}
Log_E (F)=\{g\in\Sigma_\infty(E) : \forall x \in E  (x\ge g \Rightarrow x \in E^F )\}
\end{equation}
Since ordinary words are strings, $F$ can be regarded as a set of strings, hence $E^F$ is defined. $E^F$ is the relative cylindrification of $F$ in $E$, and this in turn is the set of all words in $E$ that are prefixed by words in $F$.

$Remark$ If $F$ is a cylinder in $E$, i.e., $F=E^H$ for some $H\subseteq \Sigma_\infty(E)$, then $E^F=F$ by Equation~\ref{mauri6}.

$Remark$ For $E=\Sigma^*$ Equation~\ref{mauri7} gives $absolute$ $logogram$.

The main property of correspondence $Log_E : E \rightarrow \Sigma_\infty(E)$ is the following.  Given $A, B \subseteq E$,
\begin{equation}\label{mauri8}
Log_E (A \cup B) \supseteq Log_E(A) \cup Log_E(B).
\end{equation}
Let us understand this inclusion. $Log_E (A \cup B)$ is the set of all strings that, for $x$ in $E$, are able to trigger event $x\in E^{A \cup B}=E^A \cup E^B$. A string that triggers $x\in E^A$ certainly belongs to $Log_E (A \cup B)$. Analogously, a string that triggers $x\in E^B$ certainly belongs to $Log_E (A \cup B)$. Thus, $Log_E(A) \cup Log_E(B)$ certainly is a subset of $Log_E (A \cup B)$. However, there can be strings $f$ whose inclusion in a word $x \in E$ is a sufficient condition for event $x\in E^A \cup E^B$ but not for $x\in E^A$ or $x\in E^B$. Thus, in the general case $Log_E (A \cup B)$  is not the same set as $Log_E(A) \cup Log_E(B)$.

\subsection{Entanglement}
The presence of certain strings in a word may entail that of certain others. Given $H, K \subseteq \Sigma_\infty (E)$, we write $K \sqsupseteq^E H$ if the following happens: Every word in $E$ which includes strings from $K$ also includes strings from $H$. (Think of strings in $K$ as spies, or else symptoms, for presence in an input string $x$ of strings from $H$.) If $H \sqsupseteq^E K$ and $K \sqsupseteq^E H$ then we write $H \equiv^E K$ and say that $H, K$ are $isoexpansive$ relative to $E$. Clearly, $\equiv^E$ is an equivalence relation. It is easily seen that $H \equiv^E K$  if and only if  $E^H=E^K$.

For $E=\Sigma^*$ we rewrite $\sqsupseteq^E$ as $\sqsupseteq$ and $\equiv^E$ as $\equiv$. Note that $f \sqsupseteq^E g$ if and only if every word $x$ (within $E$) which includes $f$ also includes $g$.

 We mention a few easy facts. (i) If $f \le g$ then $f \sqsupseteq^E g$ for any possible $E$. (ii) In the general case $f \sqsupseteq^E g$ does not imply $f \le g$. (It is well possible that this holds for specific sets $E$. For example, if $E=\Sigma^*$ then $f \sqsupseteq^E g$ if and only if $f \le g$.) (iii) If $f, g$ are incompatible, then it cannot be that $f \sqsupseteq^E g$. (iv) Given any $H, K \subseteq \Sigma_\infty (E)$,
\begin{equation}\label{mauri11}
H \subseteq K \Rightarrow H \sqsupseteq K \Rightarrow H \sqsupseteq^E K
\end{equation}
We ask: Is there any easy piece of algebra linking expansion, logogram, entanglement? To get an answer, we define a Galois connection that will provide us with a closure operation in $\Sigma_\infty (E)$, noted $H \rightarrow H^{\alpha\beta}$  . We will see that $H$ and $H^{\alpha\beta}$ are isoexpansive relative to $E$. What more, there can be distinct subsets $K, I$,.. of $H$ being isoexpansive (mod $E$) to $H^{\alpha\beta}$ while possibly exhibiting different computational behaviors.

We define our connection to be a pair $(\alpha, \beta)$ of correspondences between sets of strings and sets of words. The first correspondence $\alpha$ carries a set of strings $H \subseteq \Sigma_\infty (E)$ into a corresponding set of words $H^{\alpha} \subseteq E$. The second carries a set of words $A \subseteq E$ into a set of strings $A^{\beta} \subseteq \Sigma_\infty (E)$ according to
\begin{equation}\label{mauri12}
H \sqsubseteq^E K \Rightarrow H^{\alpha} \supseteq K^{\alpha}
\end{equation}
\begin{equation}\label{mauri13}
A \subseteq B \Rightarrow A^{\beta} \sqsupseteq^E B^{\beta}
\end{equation}
\begin{equation}\label{mauri14}
H \sqsubseteq^E H^{\alpha \beta}, \hspace{1em} A\subseteq A^{\beta \alpha}
\end{equation}
The connection is formally defined through the explicit expressions:
\begin{equation}\label{mauri15}
H^{\alpha} = E^H
\end{equation}
\begin{equation}\label{mauri16}
A^{\beta} = Log_E (A)
\end{equation}
We emphasize that $A$ is any subset of $E$. Thus, given any subset $A$ of the reference set $E$ the function $A^{\beta} =Log_E (A)$ is defined. However, not all subsets $A$ of $E$ happen to be the conjugate set $H^{\alpha}$ of some set $H \subseteq \Sigma_\infty (E)$. If that happens, we say that $A$ is closed. Note that $A$ closed implies $E^A = A$.

\begin{theorem}
$(\alpha, \beta)$ is a Galois connection.
\end{theorem}
\begin{proof}
We must derive Equations \ref{mauri12}-\ref{mauri14} from Equations \ref{mauri15}-\ref{mauri16}.

(I) Let $H, K \subseteq \Sigma_\infty (E)$ be given, and assume $H \sqsubseteq^E K$.

Let $g \in K$, and let $x$ be any word in $E$ such that $x \ge g$. Then $x \in E^K$ hence $x \in K^{\alpha}$. Since $H \sqsubseteq^E K$, there exists $f \in H$ such that $x \ge f$. Then $x$ is in $E^H$ hence $x \in H^{\alpha}$ . Equation \ref{mauri12} is proved.

(II) Next, we prove Equation \ref{mauri13}. Let $A, B \subseteq E$ and assume $A \subseteq B$.

We must prove that if a word $x \in E$ includes a string $g$ from $A^{\beta}$  then $x$ also includes a string $f$ from $B^{\beta}$.

Let $g \in Log_E (A)$ so that $g \in A^{\beta}$  by Equation \ref{mauri16}.

Thus, for all $x \in E$ we have $x \ge g \Rightarrow x \in E^A$. But $A \subseteq B$, hence $x \in E^A \Rightarrow x \in E^B$ by Equation \ref{mauri5}.

Thus, for all $x \in E$ we have $x \ge g \Rightarrow x \in E^B$. By Equation \ref{mauri16} this is to say $g \in Log_E (B) = B^{\beta}$.
We have shown that $A^{\beta}  \subseteq B^{\beta}$. Equation \ref{mauri13} follows by virtue of Equation \ref{mauri11}.

(III) Next, we prove the first of Equations \ref{mauri14}.

Let $g \in H^{\alpha\beta}$ and let $x \in E$ be any string in $E$ such that $x \ge g$.

We have $H^{\alpha} = E^H$ hence $H^{\alpha\beta} = Log_E (E^H)$. Thus, $g \in Log_E(E^H)$. By Equation \ref{mauri7} we have $x \in Exp_E(E^H)$, and then, by virtue of Equation \ref{mauri6}, $x \in E^H$.

We conclude that there exists $f \in H$ such that $x \ge f$.

(IV) Next, we prove the second of Equations \ref{mauri14}.
It follows from Equations \ref{mauri15}, \ref{mauri16} that $A^{\beta\alpha}  = E^{Log_E (A)}$. On other hand one has $A \subseteq E^{Log_E (A)}$ for any $A \subseteq E$. Indeed, $A \subseteq E^A$ and $E^A = E^{Log_E (A)}$ from definitions, taking into account Equation \ref{mauri6}. We proved the second of Equations \ref{mauri14}.
\end{proof}
\begin{theorem}
The following equations hold:
\begin{equation}\label{mauri17}
H \subseteq H^{\alpha\beta}, \hspace{1em} A \subseteq A^{\beta\alpha}
\end{equation}
\begin{equation}\label{mauri18}
H^{\alpha} = H^{\alpha\beta \alpha}, \hspace{1em} A^{\beta} = A^{\beta \alpha \beta}
\end{equation}
\end{theorem}
\begin{proof}
From theory of Galois connection \cite{birkhoff}.
\end{proof}
\begin{theorem}
The map $H \rightarrow H^{\alpha\beta}$ is a closure operation in $\Sigma_\infty (E)$, and $A \rightarrow A^{\beta \alpha}$ is a closure operation in $E$.
\end{theorem}
Only for a closed $A$ do we have that, for all $x$ in $E$, $x \in A$ if and only if there exists a $g \in Log_E (A)$ such that $x \ge g$. If and only if $A$ is a closed subset of the reference set $E$ we define the reduced kernel $|Log_E (A)|$. Regarding the reduced kernel $|Log_E (A)|$ of a closed set $A \subseteq E$, we explicitly note that, for any $x$ in $E$, $x \in A$ if and only if there is $g \in |Log_E (A)|$ such that $x \ge g$.

\section{The Kernel of a Decision Program}
We begin with a few remarks on the nature of the strings that happen to occur in the reduced logogram $|Log_E (F)|$.
	For any NP decision problem $(E, F)$ we assume $F$ to be a relative cylinder in $E$. (It is known that $SAT$ is a cylinder \cite{balcazar}.) The strings in $|Log_E (F)|$ are certificates of membership for $F$ relative to $E$: For words in $E$, to include one or more strings from $|Log_E (F)|$ is necessary and sufficient for membership in $F$. In principle, we cannot exclude that $|Log_E (F)|$ may contain strings that behave as collective witnesses, also called wizards. (There exist problems, e.g. $PRIMES$, where $|Log_E (F)|$ has wizards.)
	In that case a program $P$ solving $(E, F)$ might do calculations that are functionally equivalent to testing input $x$ for wizards.

Let $P$ solve problem $(E, F)$. The computations that $P$ performs are functionally equivalent to sequences of tests done on input $x$. This is part of Scott's view of computations \cite{scott} \cite{larsen}. (The term ``test'' is ours: Dana Scott uses ``token'' or else ``piece of information'' according to context.)
	Note that Scott's theory is consistent with our developments as soon as we identify Scott's tokens with strings. In this view what $P$ actually does is searching the input $x$ for strings in $|Log_E (F)|$. That yields a view of computations as sequences of tests $in$ $disguise$.

Let program $P$ solve problem $(E, F)$. The tests in $|Log_E (F)|$ are those that $P$ can use: They are so to speak at disposal for a program $P$. Which of these tests are actually used by $P$ is a different story. We define the $kernel$ of program $P$, noted $Ker(P)$, to be the set of the strings from $|Log_E (F)|$ that $P$ actually uses for making decisions. The strings in $Ker(P)$ are uniquely identified by the algorithm that $P$ implements. The composition of $Ker(P)$ in terms of strings can also be determined through experiments with the executable of $P$.

A concept of great relevance for sequel is that of a $complete$ subset of the reduced logogram $|Log_E (F)|$ of decision problem $(E, F)$: We define a set $H \subseteq |Log_E (F)|$ to be complete for problem $(E, F)$ as soon as, for any $x \in E$,  one has $x \in F  \Leftrightarrow \exists f \in H  (f\le x$).

The proofs of following two theorems are not difficult and are omitted.

\begin{theorem}
A necessary condition for $P$ to correctly solve $(E, F)$ is $Ker(P)$ complete for $(E, F)$.
\end{theorem}

Let $H \subseteq |Log_E (F)|$ be a complete set of strings for $(E, F)$. We define $H$ to be $irreducible$ for $(E, F)$ as soon as no proper subset $K \subset H$ happens to be complete for $(E, F)$.

\begin{theorem}
Let  $|Log_E (F)|$ be irreducible and programs $P, Q$ both solve $(E, F)$. Then $Ker(P)=Ker(Q)$.
\end{theorem}

\section{Independence of Decision Problems}
We first introduce a notion of pairwise independence of strings relative to a reference set $E$. As next step, we define a notion of internal independence of set $E$. Next we define notions of internal and strong internal independence of a decision problem.

\paragraph{Mutual Independence of Strings}	
Let $f, g$ be any two strings in $\Sigma_\infty(E)$ where $E$ is any infinite recursive set of words over alphabet $\Sigma$. According to definitions, $f$ entangles $g$ relative to $E$ as soon as, for all $x \in E$, $x \ge f \Rightarrow x \ge g$. We agreed that $f \sqsupseteq^E g$ means that $f$ entangles $g$ relative to $E$.

Observe that $f$ fails to entangle $g$ relative to $E$ if and only if there exists $x \in E$ such that $x$ contains $f$ and does not contain $g$. If $f \not\sqsupseteq^E g$ and $g \not\sqsupseteq^E f$ then $f$ and $g$ are said $mutually$ $independent$ relative to $E$; $f$ and $g$ are $mutually$ $dependent$ relative to $E$ when they fail to be mutually independent relative to $E$. If $f, g$ are incompatible, then certainly $f, g$ are mutually independent relative to any $E$.

\paragraph{Independence of a Recursive Set} 	
Our next step is to define the internal independence of a recursive set $E$. We define $E$ to be $internally$ $independent$ as soon as, given any $f, g \in \Sigma_\infty (E)$ one has $f \sqsupseteq^E g$ if and only if $f$ is part of $g$, that is to say, if and only if $f \le g$.

\paragraph{Independence of a Decision Problem} 	
Now we are ready to introduce the simple internal independence of a decision problem $(E, F)$.
We call $(E, F)$ $internally$ $independent$ as soon as the strings in $|Log_E (F)|$ are mutually independent taken two by two.

\begin{theorem}
If $E$ is internally independent then any decision problem $(E, F)$ based on $E$ as reference set exhibits the simple internal independence property.
\end{theorem}

\begin{proof}
Let $E$ be any infinite recursive set exhibiting the internal independence property. Let $(E, F)$ be any decision problem based on $E$ as reference set. Let $f, g$ be any two strings in the reduced logogram $|Log_E (F)|$ of the problem.

(I) Assume $f, g$ incompatible. Since $f \in \Sigma_\infty (E)$, we have $E \cap Exp(f) \not= \emptyset$. Let $x \in E \cap Exp(f)$. Then $x$ is in $E$, $x$ includes $f$ and does not include $g$. Analogously, one can find a $y \in E$ which includes $g$ and does not include $f$. Thus, $f, g$ are mutually independent in $E$.

(II) Assume $f, g$ compatible. By the minimality property of the reduced logogram $|Log_E (F)|$ it cannot be that $f \ge g$. By the internal independence of the reference set $E$ one has $f \sqsupseteq^E g$ if and only if $f \ge g$. Then, it also cannot be the case that $f \sqsupseteq^E g$. As a consequence, there exists $x \in E$ which includes $f$ and does not include $g$. Analogously, there exists $y \in E$ which includes $g$ and does not include $f$. Thus, again we have that $f, g$ are mutually independent.

We conclude that problem $(E, F)$ exhibits the simple internal independence property.
\end{proof}

\paragraph{Strong Independence of a Decision Problem}
Let us now come to the strong internal independence of a decision problem. We know that $(E, F)$ is internally independent as soon as the strings in its reduced logogram $|Log_E (F)|$ are mutually independent taken two by two.

The simple internal independence of a decision problem $(E, F)$ certainly is a form of internal independence of a decision problem, but we may indeed ask for more independence: We may ask for independence of the elements of the reduced kernel $|Log_E (F)|$ taken $m$ by $m$ all $m$. The following notion of internal independence of a decision problem captures this extreme form of internal independence of a problem.
        	
We shall say that the decision problem $(E, F)$ exhibits the property of $strong$ $internal$ $independence$ if, for any choice of $s$ distinct strings $f_1,.., f_s$ in $|Log_E (F)|$, the following is true: For every $i$ between $1$ and $s$ there exists a word $x_i \in E$ such that $x_i$ contains $f_i$ and fails to contain any of the remaining strings in $\{f_1,.., f_s\}$.
	It is left for the reader to show that strong internal independence of a decision problem implies simple internal independence.

\section{Witnesses and Wizards}
From Equation~\ref{mauri8} we have
\begin{equation}\label{mauri22}
Log_E A_1 \cup .. \cup Log_E A_m \subseteq Log_E (A_1 \cup .. \cup A_m)
\end{equation}
for closed $A_1,.., A_m \subseteq E$. Now replace $m$ with $\alpha(n)$ and $A_i$ with $F_i$ :
\begin{equation}\label{mauri23}
Log_E F_1 \cup .. \cup Log_E F_{\alpha(n)} \subseteq Log_E (F_1 \cup .. \cup F_{\alpha(n)})
\end{equation}
The strings in $Log_E F_1,.., Log_E F_{\alpha(n)}$ are witnesses. The possible strings in
\begin{equation}\label{mauri24}
Log_E (F_1 \cup .. \cup F_{\alpha(n)}) - Log_E F_1 \cup .. \cup Log_E F_{\alpha(n)}
\end{equation}
we call ``wizards'' since they are so to speak able to perceive that an input $x$ shall be in someone of the $F_is$ but couldn't say which. The possible existence of this type of strings in the reduced logogram $|Log_E (F)|$ of a decision problem $(E, F)$ can be demonstrated by examples. Wizards have been found to exist in the reduced logograms of following problems (i) To decide if a symmetric loopfree graph is connected, (ii) To decide if a given positive integer is composite (note, incidentally, that $PRIMES$ is in P \cite{agrawal}).

In a situation in which the target set $F$ is decomposed according to  $ \cup_n F_n = F$, the witnesses are always there in the reduced logogram of set $F$ relative to $E$. On the contrary, the wizards may be missing. It pertains to the structure of the computational problem at hand whether the target set $F$ has wizards. We conclude this section proving a theorem:

\begin{theorem}\label{completeSubset}
If $F=\cup_n F_n$ where the $F_is$ are cylinders in $E$, then
$\cup_{i=1}^{\alpha(n)} |Log_E F_i|$ is complete for $(E, F)$.
\end{theorem}
\begin{proof}
Being a union of cylinders in $E$, $F$ is a cylinder in $E$.
Being cylinders in $E$, the $F_is$ are endowed with reduced logograms. This is to say that, for $i=1,..,\alpha(n)$ and any $x \in E$, one has $x \in F_i$ if and only if there is $g \in |Log_E (F_i)|$ such that $x \ge g$. Since the target set  $\cup_n F_n = F$ is itself a cylinder in $E$, Equation~\ref{mauri23} holds.

(I) Let $f \in |Log_E F_1| \cup .. \cup |Log_E F_{\alpha(n)}|$ and let $x$ be an input word of length $n$ such that $x \in E$ and $x \ge f$. We must prove $x \in F$.

Very obviously we have $f \in Log_E F_1 \cup .. \cup Log_E F_{\alpha(n)} $.

Since sequence $F_i$ has a cardinality function $\alpha(n)$, then, for input words $x \in E$ of length $n$, equation  $\cup_n F_n = F$ can be rewritten $F = F_1 \cup .. \cup F_{\alpha(n)}$.

Given $F_1 \cup .. \cup F_{\alpha(n)} = F$, $f \in Log_E (F)$ follows from Equation~\ref{mauri23}. Then $x \in F$ follows from $x \ge f$ (taking into account that $F$ is a cylinder in $E$).

(II) Let $x \in E$ be any input word of length $|x|=n$. Assume $x \in F$. We must prove that there exists $f \in |Log_E F_1| \cup .. \cup |Log_E F_{\alpha(n)}|$  such that $x \ge f$.
	
Since $x \in F$ and $F = F_1 \cup .. \cup F_{\alpha(n)}$, there exists $i$, $1 \le i \le \alpha(n)$, such that $x \in F_i$.

Since $F_i$ is a cylinder in $E$, the reduced logogram $|Log_E (F_i)|$ exists. This implies that, if $y \in E$ includes a string $g \in |Log_E (F_i)|$ then certainly $y \in F_i$. Conversely, if $y \in F_i$ then $y$ includes at least a string $g \in |Log_E (F_i)|$.
But this is just to say that $|Log_E (F_i)|$ is a complete subset of $|Log_E (F)|$ for $F_i$ relative to $E$, which is to say, for problem $(E, F_i)$.

Since $|Log_E (F_i)|$ is complete for $F_i$ relative to $E$, it follows from $x \in F_i$ that there exists a string $f \in |Log_E (F_i)|$ such that $x \ge f$.

Then we also have $f \in |Log_E F_1| \cup .. \cup |Log_E F_{\alpha(n)}|$.

We have shown that, given any input word $x \in E$ such that $|x|=n$, one has $x \in F$ if and only if there exists a string $f$ in $|Log_E F_1| \cup .. \cup |Log_E F_{\alpha(n)}|$ such that $f \le x$. Thus, $|Log_E F_1| \cup .. \cup |Log_E F_{\alpha(n)}|$ is a complete subset of $|Log_E F|$ for $F$ relative to $E$.
\end{proof}

\section{Application to Boolean Formulas}
The encoding scheme that we adopt converts $CNF$ formulas into words over $\Sigma=\{0, 1, 2\}$. In what follows $E=CNF$, $F=SAT$.
	
We represent clauses over $x_1,.., x_n$ by sequences of $n$ codes from $\Sigma$. Code $0$ denotes absence of the variable, code $1$ presence without minus, code $2$ presence with minus.
E.g., clause $x_1 \vee x_3 \vee -x_4$ becomes 1012.
	
A whole formula is encoded as a sequence of clauses. We define $F^{nm}$ = satisfiable formulas with $n$ variables and $m$ clauses.
	
We introduce the sequence $y_1, y_2, ..$ of solutions, and the corresponding sequence $F_1, F_2, ..$ or recursive subsets of $F$. Here the solutions $y_i$ are value assignments. The cardinality function is  $\alpha(n)=2^n$.
	We assume that $F=SAT$ as well as the regions $F_1, F_2, ..$ are closed sets in $E=CNF$. Thus, all these sets are assumed to be relative cylinders in $E$. These assumptions correspond to known properties of $SAT$ \cite{balcazar} \cite{hemaspaandra}.

Essentially, our application consists in investigating whether $|Log_E (F)|$ might possibly contain strings not already in some of the $|Log_E (F_i)|$.
	Before we discuss the propositions that we were able to derive, let us spend a few words on the logogram of $SAT$. A string in $|Log_E (F^{nm})|$ is a prescription that a word in $F^{nm}$ may or may not be conformant with. We may represent a string in $|Log_E (F^{nm})|$ as a word of length $nm$ over $\{\flat \}\cup \Sigma$.
	Example for $n=m=3$: String $\flat \flat 11 \flat 2 \flat 2 \flat$  prescribes that first clause shall include $x_3$, second shall include $x_1$ and $-x_3$, third shall include $-x_2$. Note that strings in $|Log_E (F^{nm})|$ only prescribe either $1$ or $2$ as values (by the minimality property of reduced logogram).

\begin{theorem}
Problem $(CNF, SAT)$ exhibits the strong internal independence property.
\end{theorem}
\begin{proof}
We consider $s$ distinct strings $f_1,.., f_s$ in $|Log_E (F^{nm})|$. Thus, regarded as a partial function, each $f_i$ will assign only values $1$ or $2$.
We must prove that for each $i=1,.., s$ there exists a string $x_i \in E^{nm} = CNF^{nm}$ such that $x_i$ includes $f_i$ and does not include any of the remaining strings $f_1,.., f_s$.

Let $i$ be any one of the indices $1,.., s$. Then $Dom(f_i) \subseteq \{1,.., nm\}$ and, for all $h \in Dom(f_i)$, we either have $f_i (h)=1$ or $f_i (h)=2$.

Let $x_i$ be that word of length $nm$ over $\Sigma=\{0, 1, 2\}$ such that for all $h \in Dom(f_i)$ it holds that $x_{ih}=f_i(h)$ while for $h$ not in $Dom(f_i)$ one has $x_{ih}=0$. Then certainly $x_i$ includes $f_i$.

Let $f_j$ be any one of the strings $f_1,.., f_s$ being different from $f_i$. Thus, $f_j \not= f_i$. We must prove that $x_i$ does not include $f_j$.

(I) Assume $Dom(f_j)=Dom(f_i)$.

Since $f_i$ and $f_j$ are different, there is $k \in Dom(f_i)$ such that $f_i(k) \not= f_j(k)$.

But $x_{ik} = f_i(k)$, then $x_{ik} \not= f_j(k)$. Then $x_i$ does not include $f_j$.

(II) Let Assume $Dom(f_j) \not=Dom(f_i)$.

Then either there is $a \in Dom(f_j)$ such that $a \not\in Dom(f_i)$ or there exists $b \in Dom(f_i)$ such that $b \not\in Dom(f_j)$.

Assume that $a$ exists. Then $x_i$ does not include $f_j$ since $x_{ia}=0$ while $f_j(a) \notin 0$, hence $x_{ia} \not= f_j(a)$. Analogously, $x_i$ does not include $f_j$ in case $b$ exists.
\end{proof}
\begin{theorem}\label{nowizards}
The reduced logogram $|Log_{CNF} (SAT)|$ does not contain wizards.
\end{theorem}

\begin{proof}
We must prove:
\begin{equation}\label{mauri25}
Log_E F_1^{nm} \cup .. \cup Log_E F_{\alpha(n)}^{nm} = Log_E (F^{nm})
\end{equation}
where  $\alpha(n)$ is the cardinality function of sequence $F_1, F_2, ..$. Here $F_i$ is the range of the value assignment $y_i$ (set of formulas in $F$ that are satisfied by $y_i$) and is a cylinder in $E$. Since Equation~\ref{mauri23} holds, we just have to prove that the right-hand side of Equation~\ref{mauri25} does not contain wizards. We actually will prove:
\begin{equation}\label{mauri26}
|Log_E F_1^{nm}| \cup .. \cup |Log_E F_{\alpha(n)}^{nm}| = |Log_E (F^{nm})|
\end{equation}
which is evidently equivalent to Equation~\ref{mauri25}.

We write $K^{nm}$ for $|Log_E (F^{nm})|$ and, for every integer $i=1,.., \alpha(n)$, we write $K_i^{nm} = |Log_E (F_i^{nm})|$. We must prove $K^{nm} = K_1^{nm} \cup .. \cup K_{\alpha(n)}^{nm}$.

First of all, note that the set of all witnesses $K_1^{nm} \cup .. \cup K_{\alpha(n)}^{nm}$ is complete for the target set $F=SAT$ relative to reference set $E=CNF$ by Theorem~\ref{completeSubset}. This implies that, if $x \in F^{nm}$, then $x$ includes a string $f \in K_1^{nm} \cup .. \cup K_{\alpha(n)}^{nm}$.

Let $h \in K^{nm}$. Since $K^{nm}$ is included in $\Sigma_\infty (E)$, we have $h \in \Sigma_\infty (E)$. Then there is an $x \in E^{nm}$ such that $x \ge h$. On the other side, if $x$ is in $E^{nm}$ and includes string $h$, then $x \in F$, hence, since $K_1^{nm} \cup .. \cup K_{\alpha(n)}^{nm}$ is complete for $F$ relative to $E$, there shall exist a string $k \in K_1^{nm} \cup .. \cup K_{\alpha(n)}^{nm}$ such that $x \ge k$ (and $h, k$ shall have to be compatible to one another).

We then set $h \rightarrow k$ to mean that (i) $k$ is a member of $K_1^{nm} \cup .. \cup K_{\alpha(n)}^{nm}$, (ii) there exists $x \in E^{nm}$ such that both $h \le x$ and $k \le x$. (Thus, $h \rightarrow k$ implies that $h$ and $k$ are compatible.)

Besides, we introduce the set $U(h)=\{k|h \rightarrow k\}$ of those witnesses (members of set $K_1^{nm} \cup .. \cup K_{\alpha(n)}^{nm}$) that are related to $h$.

Now, by way of contradiction, we assume that $h$ does not belong to $U(h)$.

We then have that the elements in the set $\{h\} \cup U(h)$ are all distinct.

By the strong internal independence of $SAT$, in correspondence to each string $f \in \{h\} \cup U(h)$ there exists a word $x \in E^{nm}$ such that $f \le x$ and for no $g\in \{h\} \cup U(h)$ being distinct from $f$ one has $g \le x$.

Let $x \in E^{nm}$ be such that $x \ge h$ and for no $g \in U(h)$ one has the inclusion $g \le x$. Word $x$ is in $F=SAT$ since $x$ includes $h$ which is an element of $|Log_E (F^{nm})|$. Besides, $x$ does not contain any element from $U(h)$. But that in turn means that $x$ does not contain any strings from the witset $K_1^{nm} \cup .. \cup K_{\alpha(n)}^{nm}$. (Should $x$ include a string $k$ from $K_1^{nm} \cup .. \cup K_{\alpha(n)}^{nm}$ that would mean that both $x \ge h, x \ge k$ would hold, hence $k$ would be related with $h$ which would imply $k \in U(h)$.)

This is absurd, since $K_1^{nm} \cup .. \cup K_{\alpha(n)}^{nm}$ is complete for $SAT$ relative to $CNF$. We conclude that $h$ is a member of $U(h)$, and hence is in $K_1^{nm} \cup .. \cup K_{\alpha(n)}^{nm}$. Since we already know that $K_1^{nm} \cup .. \cup K_{\alpha(n)}^{nm}$ is a subset of $K^{nm}$, we conclude that $K^{nm} = K_1^{nm} \cup .. \cup K_{\alpha(n)}^{nm}$. Thus, $SAT$ has no wizards.
\end{proof}

\begin{theorem}\label{irreducible}
The reduced logogram $|Log_{CNF} (SAT)|$ is irreducible.
\end{theorem}
\begin{proof}
Let $g \in |Log_E (F^{nm})|$. By Theorem~\ref{nowizards} we know that $g$ must be a witness. Thus, $g$ is a string conveying the specification of exactly one value assignment. Besides, $g$ is minimal (no proper restriction of $g$ is a sufficient condition for event $x \in F$). These two facts make it a straightforward task to specify the general shape that string $g$ shall exhibit.

First of all, $Dom(g)$ shall have to be a set of exactly $m$ numbers taken from
$\{1,.., nm\}$. The first of these numbers is to be taken from the first block $\{1,.., n\}$ (where the first clause is allocated), the second is to be from the second block $\{n+1,.., 2n\}$,.., the $m$th is from the $m$th block $\{n(m-1)+1,.., nm\}$ (where the last clause is allocated). Thus, there are $nm$ possible determinations for $Dom(g)$. We know that regarded as a prescription, $g$ can only prescribe the two values $1$ and $2$. (To help intuition, string $g$ can be thought of as a sequence of flats $\flat\flat..\flat$  of length $nm$ in which some of the flats (as many as $m$) have been replaced with $1s$ or $2s$.)

With any $g$ that satisfies the above requirements we associate a formula $\gamma(g)$ as follows. We note that, regarded as a prescription, $g$ prescribes the presence of exactly one literal in each clause of a formula $x$ consisting of $m$ clauses: We then state that the $i$th clause of $\gamma(g)$ shall consist of exactly the single literal that $g$ prescribes to the $i$th clause of $x$.

Evidently,  $\gamma(g)$ is satisfiable and $g \le \gamma(g)$. We claim that $\gamma(g)$ does not include members of $|Log_E (F^{nm})|$ other than $g$.

Indeed, the strings in $|Log_E (F^{nm})|$ never prescribe 0 as value, and $g$ is the largest string being included in the codeword of $\gamma(g)$ which does not prescribe 0 as value. Thus, the only strings that do not prescribe 0 as value and happen to be included in the codeword of $\gamma(g)$ are exactly string $g$ itself and the proper restrictions of string $g$. Since $g$ is minimal, all of its proper restrictions are not members of $|Log_E (F^{nm})|$. Thus $g$ is the only string being included in the codeword of $\gamma(g)$ to be found in $|Log_E (F^{nm})|$.

Hence, $|Log_E (F^{nm})|- \{g\}$ is not complete for $F^{nm}$ relative to $E^{nm}$.
\end{proof}

\section{SAT as Search Problem}
The search version of a decision problem consists in obtaining solutions for a given instance $x$. Thus, with any NP problem $(E, F)$ we associate the following search problem: Given $x$ find a solution $y$ for $x$ or state that no such $y$ exists.

It is known that, by self-reducibility of $SAT$, if we had a polynomial algorithm for $SAT$, then we would also have a polynomial algorithm for the search problem associated with $SAT$ \cite{hemaspaandra}. The results of previous sections show that we can say more: It is impossible to solve $SAT$ without at the same time solving the search problem associated with $SAT$.

These remarks suggest that we may wish to focus on the search problem associated with $SAT$. This is what we do in this section.

Given any NP problem $(E, F)$, we introduce the $cover$ of the target set $F$ associated with $|Log_E (F)|$ to be the family of sets
\begin{equation}
\mathcal D_E (F) = \{Exp_E (g) \subseteq F : g \in |Log_E (F)|\}.
\end{equation}
Its members are the $charts$ or else $regions$ of the cover. The cover that is associated with the kernel of a program $P$ solving $(E, F)$ is then
\begin{equation}
\mathcal F_P (E, F) = \{Exp_E (g) \subseteq F : g \in Ker(P)\}.
\end{equation}
Both $\mathcal D_E (F)$ and $\mathcal F_P (E, F)$ are families of subsets of the target set $F$ whose union is $F$, with $\mathcal F_P (E, F)$ being a subfamily of $\mathcal D_E (F)$.

For $SAT$ we have the following situation: $\mathcal F_P (E, F) = \mathcal D_E (F)$ by Theorem~\ref{irreducible} and the strings in $|Log_E (F)|$ are all witnesses by Theorem~\ref{nowizards}. Thus any of these strings, call it $g$, has an associated relativized cylinder $Exp_E (g)$ being fully included in only one of the regions $F_is$.

Since for $E=CNF$, $F=SAT$, $Exp_E (g)$ is actually an intersection of two absolute cylinder sets $Exp(g)$ and $E$, then $Exp_E (g)$ itself is an absolute cylinder. In general, $Exp_E(g)$ will intersect certain other regions $F_h$, $F_k$,.. , but there exists only one region $F_j$ which completely includes $Exp_E(g)$. Besides, every region $F_i$ shall have to include at least one such elementary relativized cylinder $Exp_E(g)$.

As a consequence, the cardinality of the cover $\mathcal D_E (F)$ cannot be smaller than that of the family of sets $\{F_i^n : i=1,.., 2^n\}$, hence it is exponential.

\paragraph{Remarks on the Time Complexity of SAT}
In the rest of this section we make remarks on the time complexity of SAT in the light of Theorems 8, 9, 10. We will be less formal than in previous sections. Our remarks consist of two parts:

\paragraph{Part One}
It follows from Theorem~\ref{irreducible} that there is a unique subfamily $\mathcal F$ of $\mathcal D_E (F)$ such that $F = \bigcup \mathcal F$, namely $\mathcal D_E (F)$ itself. As a consequence, for any proper subset $\mathcal F \subset \mathcal D_E (F)$ one has $F \not= \bigcup\mathcal F$.

We then have that it cannot be that $\mathcal F_P (E, F)$ is a proper subfamily of the full cover $\mathcal D_E (F)$, otherwise we would have $F \not= \bigcup\mathcal F_P (E, F)$, and then $P$ could not be correct as a program. In particular, since $\mathcal D_E (F)$ is exponential, $\mathcal F_P (E, F)$ is not allowed to be a polynomial subfamily of $\mathcal D_E (F)$ ): No search algorithm for $SAT$ can only search a polynomial family of sets.

\paragraph{Part Two}
It remains for us to discuss the possibility that one single algorithm can solve the full search problem for $x$ by directly searching the full exponential family $\mathcal D_E (F)$ in polynomial time. However this can scarcely be the case due to complete absence of any form of dependence among subsets in the reduced logogram $|Log_E (F)|$ for $E=CNF$, $F=SAT$. By this lack of internal dependence, any computation of a program $P$ solving $(CNF, SAT)$ is such that the result of any computation step does not change the results that are left possible for the subsequent steps. In the rest of this part we make a few informal remarks on how this lack of dependence comes into play.

We take a general purpose program machine $M$ as computation model. (That $M$ is a program machine means that the process carried out by $M$ is determined by a running program.) We assume that only one program is running at any moment of time within $M$. We keep machine $M$ fixed while we consider an infinite set of programs solving $SAT$ (actually the set of all programs that run on $M$ and solve $SAT$). We emphasize that the hardware is kept fixed while different programs all running on that hardware are compared.

Let $B(x, m)$ be a program which for any given input $x$ of size $n$ and every integer $m$ between $1$ and $2^n$ will decide if $x$ has solutions in the range between $y_1$ and $y_m$. Take $Time_B(x, m)$ be the number of time units that $B$ uses on inputs $x, m$.

We will make remarks that convey evidence for following statement: If for any $x$ and $m<2^n$ we have $Time_B (x, m) = Time_B (x, m+1)$, then we may replace $B$ with a new program $C$ running on $M$ and such that $Time_C (x, m) < Time_C (x, m+1) = Time_B (x, m+1)$.

Indeed, under the above hypotheses on $M$, we can speak of the the class of all programs $B$, $C$,.. that solve $SAT$ on machine $M$, and we can introduce a most efficient program $A$ in this class. We understand that $A$ is a most efficient program as soon as $Time_A(x, 2^n) \le Time_C(x, 2^n)$ for any other program $C$ on any input word $x$.

It is sufficient for us to give a hint for $Time_A(x, 1) < Time_A(x, 2)$.

Our hint is the following. Since, by Theorem~\ref{nowizards}, we have $|Log_E (F_1 \cup F_2)| = |Log_E (F_1)| \cup |Log_E ( F_2)|$ and $|Log_E (F_1)| \cap |Log_E ( F_2)|=\emptyset$, a computation that implements the collection of tests in $|Log_E (F_1 \cup F_2)|$ consists of two distinct computations, one implementing collection $|Log_E (F_1)|$ and the other implementing collection $|Log_E (F_2)|$. Thus, computation $A(x, 1)$ being a proper prefix of computation $A(x, 2)$ is compatible with assumed optimality of $A$, whence $Time_A(x, 1) < Time_A(x, 2)$.

\section{On Ascribing Knowledge to Programs}
Our theory has roots in the body of formalized concepts referred to as Scott's theory of computation \cite{dizenzo}. Thus, the reduced logogram $|Log_E (F)|$ associated with problem $(E, F)$ is an $information$ $system$ \cite{scott}, \cite{larsen} (however, the very important relation is not entailment but entanglement). Even more relevant are the relationships with the ``dynamical'' part of Scott's theory, the one regarding computations as sequences of steps through which the running program's knowledge increases \cite{gierz}. We also, in this latter respect, used concepts from the model theoretic analysis of program knowledge \cite{fagin}.

In this section we briefly review relationships of the above theory with formalisms that ascribe knowledge to a running program.

 In Scott's theory the computations that program $P$ does are functionally equivalent to sequences of tokens (or tests) being consistent with the input string $x$. In our developments, the ``tests'' or ``tokens'' are identified with the strings in $Ker(P)$. In Scott's theory, the state of knowledge of a running program $P$ consists of a pile of $assertions$. These are consistent (indeed, they are propositions that are true of one and the same object $x$). As soon as the pile becomes a decisive one, the program makes its decision and stops. Our addition is: The ``assertions'' are of the form $x \in Exp(g)$ or else $x \not \in Exp(f)$ where $f, g \in Ker(P)$.

 Searching $x$ for a string $g$ amounts to same as asking if $x$ happens to belong in the absolute elementary cylinder $Exp(g)$ associated with $g$. We thus arrive at the conclusion that all that $P$ can possibly do to make a decision consists in asking questions of this form. Thus, the computations that $P$ performs are just sequences of tests $in$ $disguise$. Note that $P$ has not got to ask whether $x$ is in $Exp_E(g)$ since $P$ already knows that $x$ is in $E$. (This is an important point since asking if $x$ is in $Exp_E(g)$ would be more computationally expensive.)

 In this theory, information regarding $x$ is acquired by $P$ in lumps. The acquisition of a piece of information occurs at the moment when the execution of a sequence of tests is completed (i.e., when the computation that implements that sequence of tests is completed). We may well think of a piece of information as being a piece of paper carrying a written note such as ``$x$ is in $Exp(g)$'' or ``$x$ fails to be in $Exp(g)$.'' These notes stack one upon the other until the pile becomes a decisive one: This is the case when the data that was gathered entails one of the events $x \in F$ or else $x \in E-F$.
	
Note that loading an input $x$ in memory does not imply computations, hence no tests are made on $x$ while loading, hence no knowledge is acquired about $x$. After loading $x$, the pile of assertions that represents program's knowledge is empty.

\section{Conclusions}
We advocated strings (with special meaning for the term) as a fundamental notion for studies of computation. So to speak, strings are needed to express the notions of internal and strong internal independence of a decision problem that underly our theory of decision problems. We were led to formulate strings to become able to derive the very basic notion of internal independence of a decision problem. Strings seem to be useful since they are absolutely elementary. Note that they are already at work in Computability. The ``restrictions'' that are often used in the study of circuit complexity are finite Boolean versions of the strings \cite{hemaspaandra}.
       		
Strings are not made of consecutive letters. A string can be interspersed in a word: By canceling zero or more letters in a word $x$, and by leaving blanks in places of letters, we get a string $f$ which is a substring of the original word $x$. In a string, one has information associated with spaces between letters (and hence with possible multiple periodicity with which letters may occur). As soon as we have the strings, we are able to define the kernel $Ker(P)$ of a decision program $P$, a set of strings which capture structural features of both program $P$ and the decision problem $(E, F)$ that $P$ solves.

 $Ker(P)$ is a subset of the reduced logogram $|Log_E(F)|$ of target set $F$ in base $E$. The reduced logogram consists of substrings of the words in $F$ which exhibit the following property: If a word in $E$ includes one of these substrings then it belongs to $F$. We may think of the strings in $|Log_E(F)|$ as kind of genes of the words in $F$. (In early notes the logogram was the $jinnee$ or $genie$ of problem $(E, F)$.) The idea clearly comes from biology, where it is known that certain occurrences at given intervals of certain letters within DNA sequences convey structural information, and yield observable characters in the macroscopic development of the structures.
	
Our application to $SAT$ uses a structural property of that problem that seems to have escaped attention so far. We called it ``strong internal independence.'' Theorem 8 shows that $SAT$ exhibits the strong internal independence property. Theorem 9 shows that, by that property, $SAT$ cannot have collective certificates in its reduced logogram. As a consequence, all the programs that solve $SAT$ have same kernel (Theorem 10).

The remarks in Section 8 suggest how Theorems 8, 9, 10 can possibly be used to put SAT under scrutiny. Our ultimate concern in this paper has been to set forth our developments as a possible new technique to attack decision problems, where ``technique'' is here used in the sense that Hemaspaandra and Ogihara gave to this term in the preface of their ``Companion.''

\section{Acknowledgements}
In the development of this research I received advice from Proff.  Fabrizio Luccio, Johan Hastad, Giancarlo Mauri, and Claudio Procesi. These results would not have been achieved without that help.

\end{document}